# Semi-quantum key distribution with single photons in both polarization and spatial-mode degrees of freedom

Tian-Yu Ye*, Hong-Kun Li, Jia-Li Hu

College of Information & Electronic Engineering, Zhejiang Gongshang University, Hangzhou 310018, P.R.China

**Abstract:** In this paper, a novel semi-quantum key distribution (SQKD) protocol is designed based on single photons in both polarization and spatial-mode degrees of freedom, which allows to establish a raw key between one quantum communicant and one classical communicant. The proposed SQKD protocol only adopts one kind of quantum state as the initial quantum resource. The detailed security analysis shows that it can resist Eve's active attacks, such as the intercept-resend attack, the measure-resend attack, the Trojan horse attack and the entangle-measure attack. The proposed SQKD protocol only needs single photons in both polarization and spatial-mode degrees of freedom as quantum resource and employs single-photon measurements. Thus, it has excellent feasibility, since the preparation and the measurement of a single photon in both polarization and spatial-mode degrees of freedom can be easily acheived with present quantum technologies.
**Keywords:** Quantum cryptography; semi-quantum key distribution (SQKD); single photon; polarization degree of freedom; spatial-mode degree of freedom

## 1  Introduction

In 2007, Boyer et al. [1] constructed the first semi-quantum key distribution (SQKD) protocol, where the novel concept of semi-quantumness was also put forward for the first time. In a SQKD protocol, a raw key can be established between one quantum communicant and one classical communicant. The classical communicant is restricted within the following operations [1-2]: (a) measuring the qubits in the fixed orthogonal basis $\{|0\rangle,|1\rangle\}$; (b) preparing the (fresh) qubits in the fixed orthogonal basis $\{|0\rangle,|1\rangle\}$; (c) sending or returning the qubits without disturbance; and (d) reordering the qubits (via different delay lines). Here, the fixed orthogonal basis $\{|0\rangle,|1\rangle\}$ is thought to be classical, as it does not refer to any quantum superposition state. Afterward, numerous good SQKD protocols [3-15] have been designed from different viewpoints. For example, in 2009, Zou et al. [3] proposed SQKD protocols using less than four quantum states; Zhang et al. [4] suggested a novel SQKD network protocol. In 2015, Zou et al. [8] designed a SQKD protocol without invoking the classical communicant's measurement capability; Krawec [9] put forward a mediated SQKD protocol. In 2016, Krawec [10] proposed a SQKD protocol where reflections contribute to the secret key. In 2018, Zhang et al. [12] proposed a single-state SQKD protocol; Tsai and Hwang [13] put forward a SQKD protocol robust against combined collective noise. However, in 2019, Tsai and Yang [14] pointed out that the protocol of Ref.[13] violates the definition of a semi-quantum environment and suggested an improvement accordingly; Wang et al. [15] constructed an efficient SQKD protocol without entanglement.

Recently, different from those protocols only working with the polarization states of photons, several quantum cryptography protocols working with single photons in both polarization and spatial-mode degrees of freedom [16-19] have been constructed. In a quantum communication protocol, the capacity of quantum communication may be automatically improved if the single photons in two degrees of freedom are used to replace those in one degree of freedom. In this paper, in order to improve the capacity of quantum communication, we are devoted to designing an SQKD protocol with single photons in both polarization and spatial-mode degrees of freedom. The rest of this paper is arranged as follows: the proposed SQKD protocol is depicted in Sect.2; its security is validated in Sect.3; and finally, discussion and conclusion are given in Sect.4.

## 2  Protocol description

A single-photon state in both the polarization and the spatial-mode degrees of freedom can be described as [16]

$$|\phi\rangle = |\phi\rangle_P \otimes |\phi\rangle_S , \qquad (1)$$

---

*Corresponding author:
 E-mail：happyyty@aliyun.com

where $|\phi\rangle_P$ and $|\phi\rangle_S$ are the single-photon states in the polarization and the spatial-mode degrees of freedom, respectively. $Z_P = \{|H\rangle, |V\rangle\}$ and $X_P = \{|S\rangle, |A\rangle\}$ are the two nonorthogonal measuring bases in the polarization degree of freedom, respectively. Here,

$$|S\rangle = \frac{1}{\sqrt{2}}(|H\rangle + |V\rangle), \quad |A\rangle = \frac{1}{\sqrt{2}}(|H\rangle - |V\rangle), \qquad (2)$$

where $|H\rangle$ and $|V\rangle$ denote the horizontal and vertical polarizations of photons, respectively. $Z_S = \{|b_1\rangle, |b_2\rangle\}$ and $X_S = \{|s\rangle, |a\rangle\}$ are the two nonorthogonal measuring bases in the spatial-mode degree of freedom, respectively. Here,

$$|s\rangle = \frac{1}{\sqrt{2}}(|b_1\rangle + |b_2\rangle), \quad |a\rangle = \frac{1}{\sqrt{2}}(|b_1\rangle - |b_2\rangle), \qquad (3)$$

where $|b_1\rangle$ and $|b_2\rangle$ denote the upper spatial mode and the lower spatial mode of photons, respectively.

The single-photon state $|\phi\rangle = |\phi\rangle_P \otimes |\phi\rangle_S$ can be produced with a 50:50 beam splitter (BS) in principle. Concretely speaking, a sequence of single-photon polarization state $|\phi\rangle_P$ is first generated, and the spatial-mode states $|\phi\rangle_S$ are prepared by BS, which is shown in Fig.1. What the BS does is to accomplish the transformations of the single-photon state in spatial-mode degree of freedom [16]. The quantum states in the polarization degree of freedom and those in the spatial-mode degree of freedom are commutative, so they can be operated independently [16].

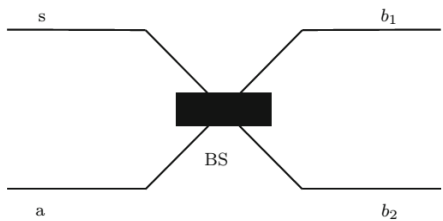

Fig.1  Schematic diagram of a Hadamard operation on a spatial quantum state of a single photon with beam splitter

There are two communicants, Alice and Bob, who want to establish a shared key between them through the transmission of quantum signals. Here, Alice is assumed to have quantum capabilities, while Bob only has classical capabilities. They agree on beforehand that $|H\rangle \otimes |b_1\rangle, |H\rangle \otimes |b_2\rangle, |V\rangle \otimes |b_1\rangle$ and $|V\rangle \otimes |b_2\rangle$ denote the classical bits 00, 01, 10 and 11, respectively.

Inspired by Zou et al.'s one-state SQKD protocol [3] and Liu et al.'s QSDC protocol [16], we put forward an SQKD protocol with single photons in both polarization and spatial-mode degrees of freedom as follows.

**Step 1:** Alice prepares $N = 4L(1+\delta)$ single photons in both polarization and spatial-mode degrees of freedom all in the state of $|S\rangle \otimes |s\rangle$, where $\delta$ is a fixed parameter greater than 0. Then, Alice sends these single photons to Bob.

**Step 2:** When single photons arrive, Bob chooses randomly either to directly reflect each of them back to Alice (this action is called as CTRL), or to measure each of them with the base $Z_P \otimes Z_S$ and resend it to Alice in the same state he found (this action is called as SIFT).

**Step 3:** Alice informs Bob of her receipt of single photons and stores them in quantum memory. Bob publishes the positions of single photons which he chose to CRTL.

**Step 4:** Alice measures each of the single photons which Bob chose to CRTL in the base $X_P \otimes X_S$ and each of the single photons which Bob chose to SIFT in the base $Z_P \otimes Z_S$.

**Step 5:** They checks the error rate on the single photons which Bob chose to CRTL. If the error rate is abnormally high, the communication will be abondoned; otherwise, it will be continued.

**Step 6:** Alice randomly chooses $L$ single photons which Bob chose to SIFT and checks the error rate on them with Bob. If the error rate is abnormally high, the communication will be abondoned; otherwise, it will be continued.

**Step 7:** Alice and Bob selects the first $L$ remaining measurement results of single photons which Bob chose to SIFT to produce the raw key bits.

For clarity, the raw key bit derivation rule of the above protocol is summarized in Table 1.

Table 1  Raw key bit derivation rule

| The initial state of single photon prepared by Alice | Bob's choice | The state received by Alice | The raw key bit shared between Alice and Bob |
|---|---|---|---|
| $\|S\rangle \otimes \|s\rangle$ | CTRL | $\|S\rangle \otimes \|s\rangle$ | None |
| | SIFT | $\|H\rangle \otimes \|b_1\rangle$ | 00 |
| | | $\|H\rangle \otimes \|b_2\rangle$ | 01 |
| | | $\|V\rangle \otimes \|b_1\rangle$ | 10 |
| | | $\|V\rangle \otimes \|b_2\rangle$ | 11 |

## 3  Security analysis

In order to get something useful about the raw key bits, an outside eavesdropper, Eve, may try to launch some attacks on the transmitted single photons, such as the intercept-resend attack, the measure-resend attack, the Trojan horse attacks and the entangle-measure attack.

(1) The intercept-resend attack

Eve may launch the intercept-resend attack in the following manner: she prepares her fake single photons beforehand, intercepts the single photons sent from Alice to Bob and transmitts her fake ones instead of the genuine ones to Bob. With respect to this attack, Eve will be discovered inevitably due to two aspects: on the one hand, Eve has to randomly prepare her fake single photons; on the other hand, Bob's operations are random to her. Without loss of generality, take Eve's fake single photon in the state of $|H\rangle \otimes |s\rangle$ for example. Eve sends $|H\rangle \otimes |s\rangle$ to Bob after intercepting the genuine single photon from Alice. If Bob chooses the CTRL mode, Alice will receive $|H\rangle \otimes |s\rangle$ and use the base $X_P \otimes X_S$ to measure it. As a result, after Alice's measurement, the state $|H\rangle \otimes |s\rangle$ is collapsed into $|S\rangle \otimes |s\rangle$ or $|A\rangle \otimes |s\rangle$ each with the probability of 50%. Thus, in this case, Eve will be detected with the probability of 50% by the security check of Step 5. If Bob chooses the SIFT mode, after Bob's measurement, the state $|H\rangle \otimes |s\rangle$ will be collapsed into $|H\rangle \otimes |b_1\rangle$ or $|H\rangle \otimes |b_2\rangle$ each with the probability of 50%. Then, Alice uses the base $Z_P \otimes Z_S$ to measure the state from Bob and obtains $|H\rangle \otimes |b_1\rangle$ or $|H\rangle \otimes |b_2\rangle$ each with the probability of 50%. However, if no attack from Eve takes place, when Bob measures the genuine state $|S\rangle \otimes |s\rangle$ in the SIFT mode, after her measurement, Alice can obtain $|H\rangle \otimes |b_1\rangle$ or $|H\rangle \otimes |b_2\rangle$ or $|V\rangle \otimes |b_1\rangle$ or $|V\rangle \otimes |b_2\rangle$ each with the probability of 25%. Thus, in this case, Eve can be detected with the probability of 50% when this state is chosen for the security check in Step 6.

It should be pointed out that if Eve happens to prepare all single photons in the state of $|S\rangle \otimes |s\rangle$, she will not be discovered, as her fake single photons are all identical to the genuine ones. However, this situation takes place only with the probability of $\left(\dfrac{1}{16}\right)^N$. If $N$ is large enough, it will naturally converge to 0. To say the least, even if this situation happens, Eve still obtains nothing about the raw key bits.

(2) The measure-resend attack

Eve may launch the measure-resend attack in the following manner: Eve intercepts the single photons sent from Alice to Bob, measures them randomly in one of the four bases $\{Z_P \otimes Z_S, Z_P \otimes X_S, X_P \otimes Z_S, X_P \otimes X_S\}$ and sends the measured states to Bob. With respect to this attack, Eve will be discovered inevitably due to two aspects: on the one hand, Eve's measurements may destroy the states of single photons; on the other hand, Bob's operations are random to Eve. Without loss of generality, assume that Eve uses the base $Z_P \otimes X_S$ to measure the single photon $|S\rangle \otimes |s\rangle$. Accordingly, the single photon $|S\rangle \otimes |s\rangle$ is collapsed into $|H\rangle \otimes |s\rangle$ or $|V\rangle \otimes |s\rangle$ each with the probability of $50\%$. No matter what state the single photon $|S\rangle \otimes |s\rangle$ is collapsed into, if Bob chooses the CTRL mode, Eve will be detected with the probability of $50\%$ by the security check of Step 5; and if Bob chooses the SIFT mode, Eve will also be detected with the probability of $50\%$ when this state is chosen for the security check in Step 6.

It should be pointed out that if Eve happens to measure all single photons in the base $X_P \otimes X_S$, she will not be discovered, as her attack does not change the original states of all single photons. However, this situation takes place only with the probability of $\left(\frac{1}{4}\right)^N$. If $N$ is large enough, it will naturally converge to 0. To say the least, even if this situation happens, Eve still obtains nothing about the raw key bits.

(3) The Trojan horse attacks

During the whole communication process, there are single photons transmitted in a round trip. Eve may utilize the round particle transmission manner to launch the Trojan horse attacks, including the invisible photon eavesdropping attack [20] and the delay-photon Trojan horse attack [21-22], to obtain something useful about the raw key bits. To defeat the invisible photon eavesdropping attack, Bob can insert a filter in front of his devices to filter out the photon signal with an illegitimate wavelength before dealing with it [22-23]. To resist the delay-photon Trojan horse attack, Bob can employ a photon number splitter (PNS) to split each sample quantum signal into two pieces and use proper measuring bases to measure the signals after the PNS [22-23]. If the multiphoton rate is abnormally high, Eve's attack behavior will be discovered.

(4) The entangle-measure attack

Eve's entangle-measure attack can be modeled as two unitaries: $\hat{U}_E$ attacking particles as they go from Alice to Bob and $\hat{U}_F$ attacking particles as they go back from Bob to Alice, where $\hat{U}_E$ and $\hat{U}_F$ share a common probe space with initial state $|\varepsilon\rangle$. As pointed out in Refs.[1-2], the shared probe allows Eve to make the attack on the returning particles depend on knowledge acquired by $\hat{U}_E$ (if Eve does not take advantage of this fact, then the "shared probe" can simply be the composite system comprised of two independent probes). Any attack where Eve would make $\hat{U}_F$ depend on a measurement made after applying $\hat{U}_E$ can be implemented by unitaries $\hat{U}_E$ and $\hat{U}_F$ with controlled gates. Eve's entangle-measure attack within the implementation of the protocol is depicted in Fig.2.

**Theorem 1.** *Suppose that Eve performs attack $(\hat{U}_E, \hat{U}_F)$ on the particles from Alice to Bob and back to Alice. For this attack inducing no error in Steps 5 and 6, the final state of Eve's probe should be independent of Bob's measurement result. As a result, Eve gets no information on the raw key bits.*

**Proof.** Before Eve's attack, the global state of the composite system composed by the single photon $|S\rangle \otimes |s\rangle$ and Eve's auxiliary particle $|\varepsilon\rangle$ is $(|S\rangle \otimes |s\rangle) \otimes |\varepsilon\rangle$. After Eve has performed $\hat{U}_E$, the global state evolves into

$$\hat{U}_E (|S\rangle \otimes |s\rangle) \otimes |\varepsilon\rangle = |Hb_1\rangle |\varepsilon_{Hb_1}\rangle + |Hb_2\rangle |\varepsilon_{Hb_2}\rangle + |Vb_1\rangle |\varepsilon_{Vb_1}\rangle + |Vb_2\rangle |\varepsilon_{Vb_2}\rangle, \tag{4}$$

where $|\varepsilon_{Hb_1}\rangle, |\varepsilon_{Hb_2}\rangle, |\varepsilon_{Vb_1}\rangle$ and $|\varepsilon_{Vb_2}\rangle$ are un-normalized states of Eve's probe.

When Bob receives the state sent from Alice, he chooses either to CTRL or to SIFT. Afterward, Eve performs $\hat{U}_F$ on the state sent back to Alice.

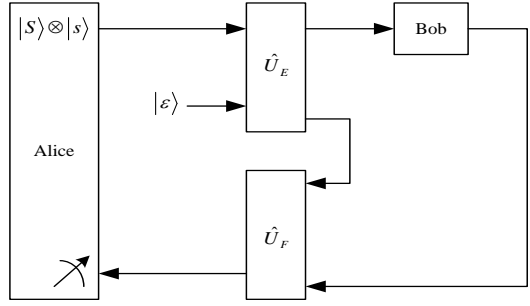

Fig.2  Eve's entangle-measure attack with two unitaries $\hat{U}_E$ and $\hat{U}_F$

(i) Firstly, consider the case that Bob has chosen to SIFT. As a result, the global state will be collapsed into either of $|Hb_1\rangle|\varepsilon_{Hb_1}\rangle$, $|Hb_2\rangle|\varepsilon_{Hb_2}\rangle$, $|Vb_1\rangle|\varepsilon_{Vb_1}\rangle$ and $|Vb_2\rangle|\varepsilon_{Vb_2}\rangle$. For Eve not being detectable in Step 6, $\hat{U}_F$ should establish the following relations:

$$\hat{U}_F(|Hb_1\rangle|\varepsilon_{Hb_1}\rangle) = |Hb_1\rangle|\zeta_{Hb_1}\rangle, \tag{5}$$

$$\hat{U}_F(|Hb_2\rangle|\varepsilon_{Hb_2}\rangle) = |Hb_2\rangle|\zeta_{Hb_2}\rangle, \tag{6}$$

$$\hat{U}_F(|Vb_1\rangle|\varepsilon_{Vb_1}\rangle) = |Vb_1\rangle|\zeta_{Vb_1}\rangle, \tag{7}$$

$$\hat{U}_F(|Vb_2\rangle|\varepsilon_{Vb_2}\rangle) = |Vb_2\rangle|\zeta_{Vb_2}\rangle, \tag{8}$$

which mean that $\hat{U}_F$ cannot change the state of single photon after Bob's measurement operation. Otherwise, Eve will be detected with a non-zero probability.

(ii) Secondly, consider the case that Bob has chosen to CRTL. As a result, the global state after Bob's operation will be kept unchanged.

After Eve performs $\hat{U}_F$ on the state sent back to Alice, due to Eqs.(5-8), the global state evolves into

$$\hat{U}_F(|Hb_1\rangle|\varepsilon_{Hb_1}\rangle + |Hb_2\rangle|\varepsilon_{Hb_2}\rangle + |Vb_1\rangle|\varepsilon_{Vb_1}\rangle + |Vb_2\rangle|\varepsilon_{Vb_2}\rangle) = |Hb_1\rangle|\zeta_{Hb_1}\rangle + |Hb_2\rangle|\zeta_{Hb_2}\rangle + |Vb_1\rangle|\zeta_{Vb_1}\rangle + |Vb_2\rangle|\zeta_{Vb_2}\rangle. \tag{9}$$

For Eve not being detectable in Step 5, Alice should measure the global state in the result $|S\rangle\otimes|s\rangle$. Consequently, it can be derived from Eq.(9) that

$$|\zeta_{Hb_1}\rangle = |\zeta_{Hb_2}\rangle = |\zeta_{Vb_1}\rangle = |\zeta_{Vb_2}\rangle = |\zeta\rangle. \tag{10}$$

(iii) Applying Eq.(10) into Eqs.(5-8) produces

$$\hat{U}_F(|Hb_1\rangle|\varepsilon_{Hb_1}\rangle) = |Hb_1\rangle|\zeta\rangle, \tag{11}$$

$$\hat{U}_F(|Hb_2\rangle|\varepsilon_{Hb_2}\rangle) = |Hb_2\rangle|\zeta\rangle, \tag{12}$$

$$\hat{U}_F(|Vb_1\rangle|\varepsilon_{Vb_1}\rangle) = |Vb_1\rangle|\zeta\rangle, \tag{13}$$

$$\hat{U}_F(|Vb_2\rangle|\varepsilon_{Vb_2}\rangle) = |Vb_2\rangle|\zeta\rangle, \tag{14}$$

respectively.

According to Eqs.(11-14), it can be concluded that for Eve not inducing errors in Steps 5 and 6, the final state of Eve's probe should be independent of Bob's measurement result. Therefore, Theorem 1 stands.

It can be concluded now that the proposed SQKD protocol is secure against an outside eavesdropper.

## 4  Discussion and Conclusion

We further compare the proposed SQKD protocol with Zou et al.'s one-state SQKD protocol [3] here. In Zou et al.'s one-state SQKD protocol [3], one raw key bit can be derived from one single photon in one degree of freedom when it is chosen by Bob to SIFT, while in the proposed SQKD protocol, two raw key bits can be derived from one single photon in both polarization and spatial-mode degrees of freedom when it is chosen by Bob to SIFT. Therefore, the proposed SQKD protocol doubles the capacity of quantum communication of Zou et al.'s one-state SQKD protocol [3].

In summary, a novel SQKD protocol is proposed by merely using single photons in both polarization and spatial-mode degrees of freedom as the initial quantum resource. We validate in detail that it can overcome Eve's active attacks, such as the intercept-resend attack, the measure-resend attack, the Trojan horse attack and the entangle-measure attack. It only employs single photons in both polarization and spatial-mode degrees of freedom as quantum resource and adopts single-photon measurements. Consequently, it is practical in reality, since the preparation and the measurement of a single photon in both polarization and spatial-mode degrees of freedom can be easily realized with present quantum technologies. Compared with Zou et al.'s one-state SQKD protocol [3], it doubles the capacity of quantum communication.

**Acknowledgments**
Funding by the Natural Science Foundation of Zhejiang Province (Grant No.LY18F020007) is gratefully acknowledged.